# Reproducible Research Can Still Be Wrong: Adopting a Prevention Approach

Jeffrey T. Leek, Associate Professor of Biostatistics and Oncology, Johns Hopkins University, jtleek@gmail.com

Roger D. Peng, Associate Professor of Biostatistics, Johns Hopkins University, rdpeng@jhu.edu

Reproducibility – the ability to recompute results – and replicability – the chances other experimenters will achieve a consistent result – are two foundational characteristics of successful scientific research. Consistent findings from independent investigators are the primary means by which scientific evidence accumulates for or against an hypothesis. And yet, of late there has been a crisis of confidence among researchers worried about the rate at which studies are either reproducible or replicable. In order to maintain the integrity of science research and maintain the public's trust in science, the scientific community must ensure reproducibility and replicability by engaging in a more preventative approach that greatly expands data analysis education and routinely employs software tools.

We define reproducibility as the ability to recompute data analytic results given an observed data set and knowledge of the data analysis pipeline. The replicability of a study is the chance that an independent experiment targeting the same scientific question will produce a consistent result (1). Concerns among scientists about both have gained significant traction recently due in part to a statistical argument that suggested most published scientific results may be false positives (2). At the same time, there have been some very public failings of reproducibility across a range of disciplines from cancer genomics (3) to economics (4), and the data for many publications have not been made publicly available, raising doubts about the quality of data analyses. Popular press articles have raised questions about the reproducibility of *all* scientific research (5) and the United States Congress has convened hearings focused on the transparency of scientific research (6). The result is that much of the scientific enterprise have been called into question, putting at risk funding and hard won scientific truths.

From a computational perspective, there are three major components to a reproducible and replicable study: (a) the raw data from the experiment are available, (b) the statistical code and documentation to reproduce the analysis are available, and (c) a correct data analysis must be performed. Recent cultural shifts in genomics and other areas have had a positive impact on data and code availability. Journals are starting to require data availability as a condition for publication (7) and centralized databases such as the NCBI's GEO are being created for depositing data generated by publicly funded scientific experiments. New computational tools such as knitr, iPython notebook, LONI, and Galaxy have simplified the process of distributing reproducible data analyses.

Unfortunately, the mere reproducibility of computational results is insufficient to address the replication crisis because even a reproducible analysis can suffer from many problems—confounding from omitted variables, poor study design, missing data—that threaten the validity and useful interpretation of the results. While improving the reproducibility of research may increase the rate at which flawed analyses are uncovered, as recent high-profile examples have demonstrated, it does not change the fact that problematic research is conducted in the first place.

The key question we want to answer when seeing the results of any scientific study is "Can I trust this data analysis?" If we think of problematic data analysis as a disease, reproducibility speeds diagnosis and treatment in the form of screening and rejection of poor data analyses by referees, editors, and other scientists in the community (Figure 1).

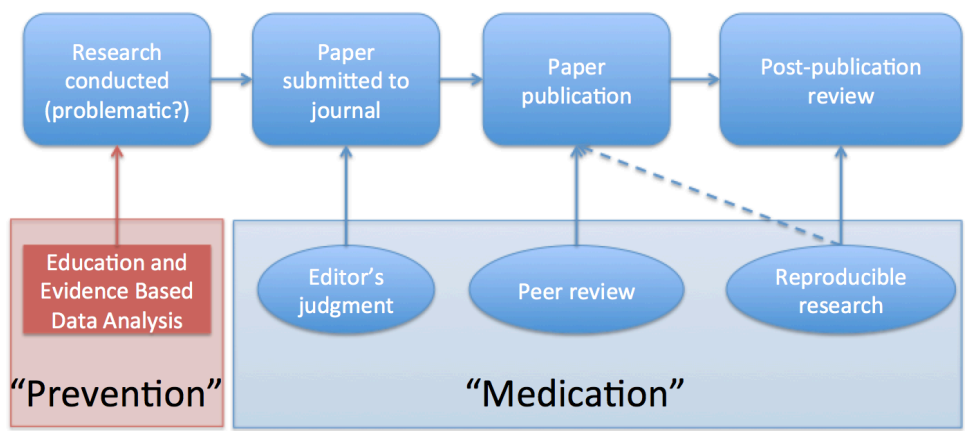

Figure 1: **Education and evidence-based data analysis as prevention.** Peer review and editor evaluation are treatments for poor data analysis. Education and evidence-based data analysis can be thought of as prevention.

This "medication" approach to research quality relies on peer reviewers and editors to make this diagnosis consistently—this is a tall order. Editors and peer reviewers at medical and scientific journals often lack the training and time to perform a proper evaluation of a data analysis. This problem is compounded by the fact that data sets and data analyses are becoming increasingly complex, the rate of submission to journals continues to increase (8), and the demands on statisticians to referee are increasing. These pressures have reduced the efficacy of peer review in identifying and correcting potential false discoveries in the medical literature. And, crucially, the medication approach does not address the problem at its source.

We suggest that the replication crisis needs to be considered from the perspective of primary prevention. If we can prevent problematic data analyses from being conducted, we can substantially reduce the burden on the community of having to evaluate an increasingly heterogeneous and complex population of studies and research findings. The best way to prevent poor data analysis in the scientific literature is to
(a) increase the number of trained data analysts in the scientific community and (b) identify statistical software and tools that can be shown to improve reproducibility and replicability of studies.

How can we dramatically scale up data science education in the short term? One approach that we have taken is through Massive Online Open Courses (MOOCs). The Johns Hopkins Data Science Specialization (http://jhudatascience.org/) is a sequence of 9 courses covering the full spectrum of data science skills from formulating quantitative questions, to cleaning data, to statistical analysis and producing reproducible reports. So far we have enrolled over 1.5 million students in this Specialization. A complementary approach is crowd-sourced short courses such as Data and Software Carpentry (http://software-carpentry.org/) that have addressed the extreme demand for data science knowledge on a smaller scale. But simply increasing data analytic literacy comes at a cost. Most scientists in these programs will receive basic to moderate training in data analysis, creating the potential for producing individuals with enough skill to perform data analysis but without enough knowledge to prevent mistakes.

To improve the global robustness of scientific data analysis, we must couple education efforts with the identification of data analytic strategies that are most reproducible and replicable in the hands of basic or intermediate data analysts. Statisticians must bring to bear their history of developing rigorous methods to the area of data science. A fundamental component of scaling up data science education is performing empirical studies to identify statistical methods, analysis protocols, and software that lead to increased replicability and reproducibility in the hands of users with basic knowledge. We call this approach *evidence-based data analysis*. Just as evidence-based medicine applies the scientific method to the practice of medicine, evidence-based data analysis applies the scientific method to the practice of data analysis. Combining massive scale education with evidence-based data analysis can allow us to quickly test data analytic practices in a population most at risk for data analytic mistakes (9).

In much the same way that epidemiologist John Snow ended a London cholera epidemic by removing a pump handle to make contaminated water unavailable we have an opportunity to attack the crisis of scientific reproducibility at its source. Dramatic increases in data science education, coupled with robust evidence-based data analysis practices, have the potential to prevent problems with reproducibility and replication before they can cause permanent damage to the credibility of science.

## References

[1] Peng RD. Reproducible research in computational science. Science (New York, Ny) 2011;334:1226.
[2] Ioannidis JP. Why most published research findings are false. PLoS medicine 2005;2:e124.
[3] Baggerly KA, Coombes KR. Deriving chemosensitivity from cell lines: Forensic bioinformatics and reproducible research in high-throughput biology. The Annals of Applied Statistics 2009;:1309–1334.
[4] Herndon T, Ash M, Pollin R. Does high public debt consistently stifle economic growth? a critique of reinhart and rogoff. Cambridge journal of economics 2014;38:257–279.
[5] Eight (no, nine!) problems with big data. http://www.nytimes.com/2014/04/07/opinion/eight-no-nine-problems-with-big-data.html?\s\do5(r)=0, 2014.
[6] Alberts B, Stodden V, Young S, Choudhury S. Testimony on scientific integrity & transparency. science.house.gov/hearing/subcommittee-research-scientific-integrity-transparency, 2013.
[7] Bloom T. PLOS's New Data Policy: Part two. http://blogs.plos.org/everyone/2014/02/24/plos-new-data-policy-public-access-data-2/, 2014.
[8] Jager LR, Leek JT. An estimate of the science-wise false discovery rate and application to the top medical literature. Biostatistics 2014;15:1–12.
[9] Fisher A, Anderson GB, Peng R, Leek J. A randomized trial in a massive online open course shows people don't know what a statistically significant relationship looks like, but they can learn. PeerJ 2014;2:e589.